# An Immune-related lncRNAs Model for Prognostic of SKCM Patients Base on Cox Regression and Coexpression Analysis

Wenjie Jiang[a,b,*], Chang Lu[a,b], Jing Qu[a,b] and Xiaoyu Mei[a,b]

a   School of Information Science and Technology, Northeast Normal University, Changchun 130117, China.
b   Institute of Computational Biology, Northeast Normal University, Changchun 130117, China.
    *Correspondence:jiangwj057@nenu.edu.cn

## Abstract

Skin Cutaneous SKCM (SKCM) is a high-risk skin cancer having an outstanding recurrence rate. The lethal rate of SKCM remains in a high level even after various therapy approaches had been applied to the patients. Currently, the prognostic is one important part of the SKCM therapy, many efforts had been made for the purpose, but it keeps as a challenging problem. In some cancer studies, the expression of lncRNA is observed to correlate with cancer, it is a promising approach to improve the prognostic therapy of SKCM.

In this study, we proposed a 4 immune-related lncRNAs model for SKCM prognosis prediction, where, data on SKCM of TCGA were used, A prognostic model for SKCM was built, separately solved Survival analysis and independent prognosis problems. The method found that the ROC score of the model reached 0.749, our method was proved to have excellent performance in prognostic, and indicating 4 immune-related lncRNAs may play a unique role in the carcinogenesis of SKCM. The highlights are the calculation method is suitable for processing the selected data and Immune-related lncRNAs were creatively selected to construct the model.

**Introduction:**

Skin cancer is the most common of all cancers. SKCM accounts for only 2 percent of skin cancer [1] but is responsible for the majority of skin cancer deaths. Advances in screening and the development of anti-cancer strategies, the treatment[2] of SKCM has improved significantly. However, SKCM still has a high recurrence rate[3]. Research shows that tumor thickness[4], the depth related to skin structures[5], the type of SKCM[6], the presence of ulceration[7], presence of lymphatic[8], presence of tumor-infiltrating lymphocytes[9], the location of lesions[10], presence of satellite lesions[11], the presence of regional[12] and other factors affect the prognosis of SKCM. Although many biomarkers have been found for the prognosis of SKCM, the pathological of SKCM[13] is complex, so the prognosis is still a difficult problem. Now, it is necessary to construct a new risk prediction model for SKCM to improve the treatment level of SKCM patients. The expression of LncRNAs is associated with cancer. Further, immune-related lncRNAs are significantly correlated with cancer[14]. So, it is necessary to establish an immune-related lncRNA model to predict SKCM survival.

In the post-genome era, many genome sequencing technologies have emerged[15]. These tools provide new ideas and insights for the diagnosis and prognosis of tumors. These next-generation sequencing methods and data can help better identify clinical biomarkers for cancer. The discovery of long non-coding RNA (lncRNA) has dramatically changed our understanding of cancer. LncRNA expression and dysregulation are more specific to cancer type than protein-coding genes[16]. Recent studies have shown that lncRNA plays an essential role in gene regulation and tumorigenesis, including proliferation, adhesion, migration, and apoptosis[17]. Considering the heterogeneity of SKCM and the complexity of non-coding RNA, a set of lncRNA biomarkers may be more accurate and stable for prognosis. Li, H et al. [18] constructed a 7-lncRNA model based on the Cancer Genome Atlas (TCGA) database, which may be able to predict the Overall Survival (OS) of breast cancer patients with high diagnostic accuracy. Liu, Z et al. [19] screened eight genes (IGHA1, IGHGP, IGKV2 28, IGLL3P, IGLV3 10, AZGP1P1, LINC00472, and SLC16A6P1) by univariate cox-regression analysis, and established a prognosis model for breast cancer from TCGA.

In this study, we screened the differentially-expressed lncRNAs associated with SKCM from the TCGA database. Then, immune-related lncRNAs were screened by co-expression and developed an immune-related lncRNA model to predict the prognosis of SKCM patients. It is well known that lncRNA can directly or indirectly affect the function of proteins and cells because they are involved in the regulation of mRNA.Therefore, we further explored the role of lncRNA in the model by studying the function of lncRNA-related mRNAs. In summary, the application of lncRNA provides a deeper understanding of the prognosis of BRCA, and that may help guide treatment.

## MATERIALS AND METHODS:

**Data Source**

Obtain the transcription data and clinical information of SKCM patients from TCGA[20]. There were 470 cases with clinical data, and 449 cases with incomplete data were excluded. A total of 472 transcriptional data samples were collected, including one healthy sample and 471 SKCM samples. The SKCM samples with incomplete prognostic information were deleted. A total of 447 SKCM samples are selected. At last, there are 360 cases for model building and 87 cases for model validation. As the data is retrieved from the TCGA database (a public database), further ethical certification does not apply to our research. Data collection and processing are in line with TCGA's data policy for personal protection.

**Identification of lncRNA and mRNA**

We from NCBI [21] Obtain the human genome annotation file (*human.gtf*) and determine the type of gene according to the *gene_biotype* item in the data. We screen the corresponding gene in transcriptome data by examining *gene_biotype* to determine its species.

**Found immune-related lncRNA and lncRNA - Related mRNA by co-expression**

Immune-related lncRNAs are closely related to cancer. To better explore the role of lncRNA in the risk assessment model, the Pearson correlation-based co-expression method[22] was used to found immune-related lncRNA. According to cor > 0.4 and P-value < 0.001, the related immune-related lncRNA was screened for building model, and the lncRNA-related mRNA were searched using the same method and criteria.

**The definition of immune-related lncRNA related prediction model**

Based on the prognostic characteristics of lncRNA, an immune-related lncRNA prognostic model was constructed. The correlation between the expression level of lncRNA and the Overall Survival rate was studied by univariate and multivariate Cox-regression analysis. Single variable Cox proportional hazard regression analysis was performed with the R-survival. The lncRNA was considered significant when the p-value was <0.001 in the univariate Cox-regression analysis, and then, the result was selected for multivariate Cox-regression analysis. Finally, Cox regression analysis was carried out to evaluate the effect of the gene as an independent prognostic factor on the survival rate of patients. The lncRNA-based prognostic risk score was used to calculate a multivariate cox-regression model based on a linear combination of regression coefficients and its expression level[23].

$$risk\ Score = \sum_{i=1}^{N} Exp_i \times \beta_i$$

LncRNA was selected by optimizing the AIC value. The survival data of 447 patients with SKCM were divided into high-risk groups and low-risk groups. Kaplan-Meier (KM) survival curve[24] was generated to evaluate the Overall Survival of low-risk or high-risk cases, and the ROC curve was analyzed to calculate AUC[25] value to evaluate the prediction ability of the model. We compared the prediction performance of the 4 immune-related lncRNAs model and traditional clinical risk factors (including age, T, N, M, stage, and er) by univariate and multivariate Cox analysis.

**Principal component analysis**

To verify the significant differences in expression between high-risk and low-risk patients, the PCA algorithm[26] was used to conduct dimensional reduction of risk model lncRNA, immune lncRNA, immune genes and all genes of patients, and then 3d visualization.

**Go and KEGG analysis of lncRNA related mRNA**

To understand the underlying biological pathways between lncRNA and the related mRNAs, we use the gene ontology database(GO) for analysis. Finally, go items with significant enrichment were selected for biological function analysis. Kyoto Encyclopedia of Genes and Genomes (KEGG). They were used for path enrichment analysis[27].

# RESULTS

**Derivation of the lncRNA prediction model**

A total of 181 differentially expressed lncRNAs need to be further studied. First of all, we conducted a univariate Cox regression analysis to study the correlation between the differential expression of lncRNA and Overall Survival of SKCM patients. A total of 10 lncRNA was obtained by P-value < 0.001, which was significantly correlated with SKCM patients. As shown in Figure1, There are ten lncRNA that meet the conditions, among which only the Hazard radio value of AC009495.2 > 1, indicating that this is a high-risk lncRNA, while the rest are low-risk lncRNAs.

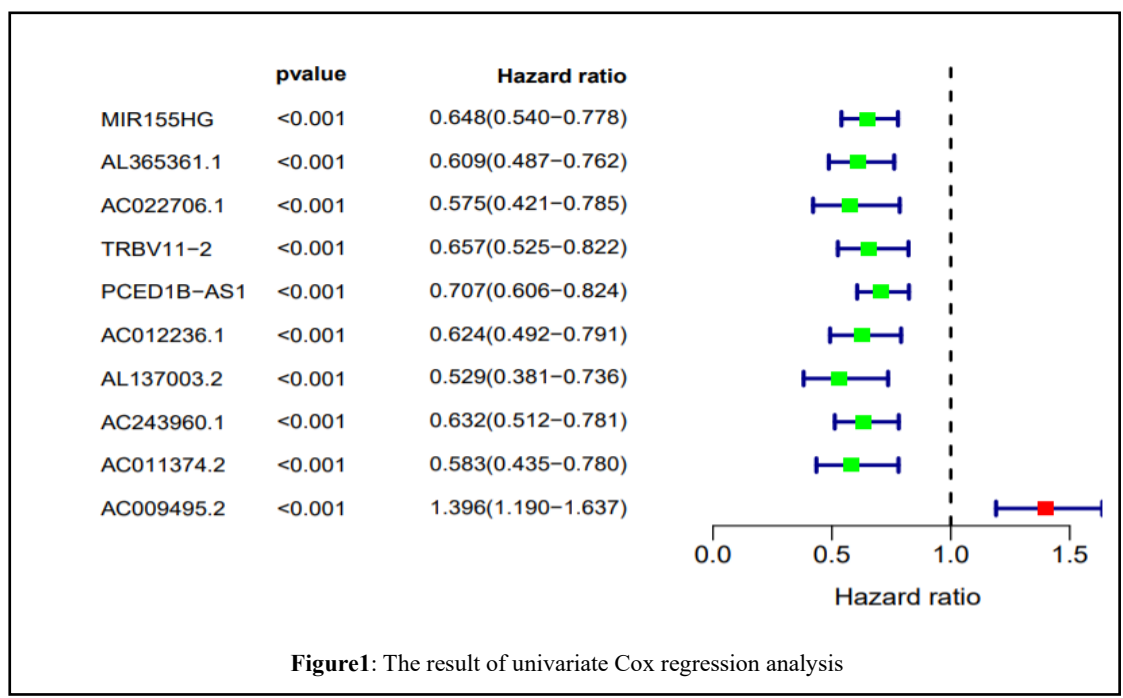

**Figure1**: The result of univariate Cox regression analysis

Then, based on preliminary screening using univariate Cox regression analysis, We obtained four lncRNAs that were used to construct the prediction model using multivariate Cox regression

analysis. RS = (-0.32396 × MIR155HG expression level) + (-0.33529 × AL137003.2 expression level) + (0.244771 × AC009495.2 expression level) + (- 0.27961 × AC011374.2 expression level).

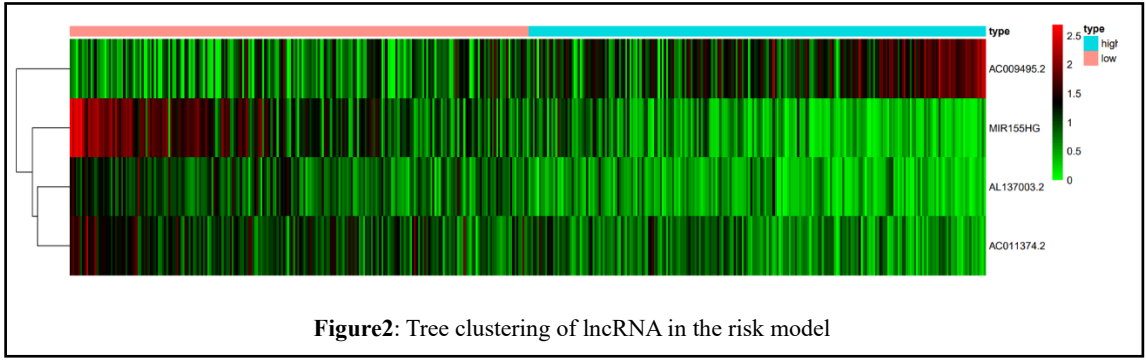

**Figure2**: Tree clustering of lncRNA in the risk model

Among the four lncRNAs obtained by Cox regression analysis(**Figure2**), AC009495.2 was positive, indicating that this lncRNA had a higher risk. Meanwhile, the other three MIR155HG、AL137003.2 and AC011374.2 were negatively correlated.

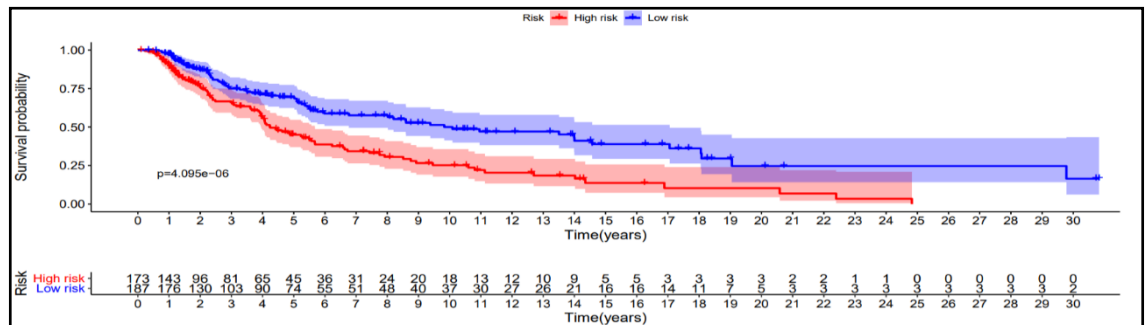

**Figure3(A)**: Kaplan-Meier survival curve analysis of the training data set

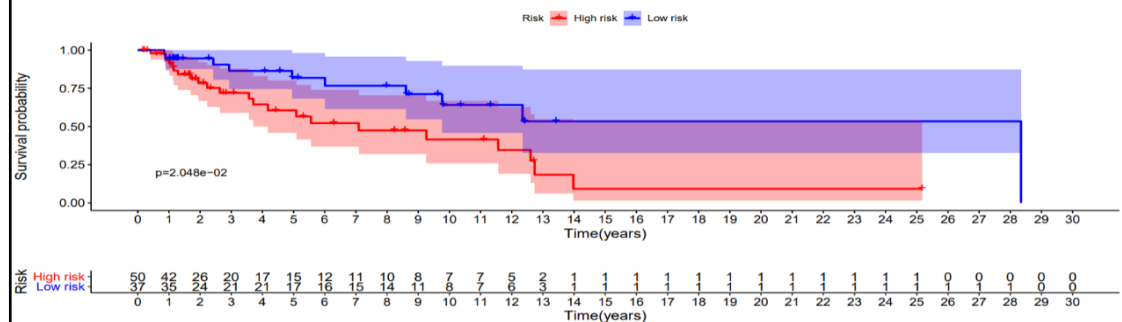

**Figure3(B)**: Kaplan-Meier survival curve analysis of testing data set

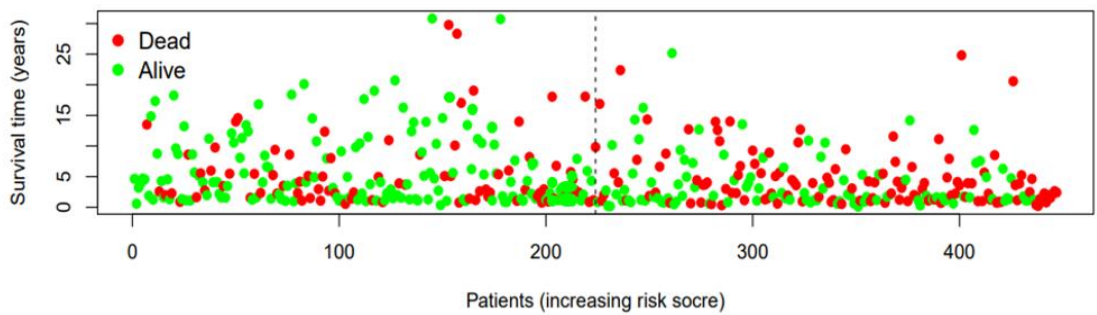

**Figure3(C)**: The patient's survival status

Although the risk associated with lncRNA is not high, they are still relevant in the prognosis model. These lncRNAs constitute the prognosis model of SKCM patients. In 447 patients with SKCM, four risk scores of lncRNA expression were calculated, and the median of prognosis score was used as the grouping threshold. With the median RS as the group threshold, 223 patients with a prognosis score higher than the RS threshold were divided into the high-risk group, and 224 patients with a prognosis score lower than the RS threshold were split into the low-risk group. According to the prognostic risk model constructed by four lncRNAs, Kaplan-Meier survival curve analysis was performed on the high-risk group and the low-risk group. It was found that the overall survival rate of the high-risk group was lower and the difference between the two groups was statistically significant(**Figure3(A&B)**). Besides, The survival rate of the training set and the test set was similar in each year in the KM curve, indicating the model is universal. We can conclude from **Figure3(C)** that the higher the risk value predicted by the 4 immune-related lncRNAs model, the more patients died. Then, the AUC of the ROC curve was calculated to evaluate the prognosis of the 4 immune-related lncRNAs model. The higher the AUC is, the better the prediction performance of the model is. The AUC of the model was 0.749 in the test data set(**Supplementary material 1**), indicating that the prediction model was highly sensitive and specific. Besides, we conducted another set of comparative experiments. Instead of using only SKCM immune-related lncRNA, we used all SKCM-related lncRNA to construct the model, and its AUC only reached 0.688.

**Clinical correlation analysis**

We plotted the ROC curve **Figure4(A)** of the clinical data and found that T, stage and N were most associated with risk. We carried out a clinical correlation analysis in the T on four lncRNA, and the results were shown in **Figure4(B)**. It was found that AL137003.2 and MIR155HG were significantly correlated with the T stage. For AL137003.2, the expression is very high at T0, and the expression level from T1 to T4 is roughly the same. For MIR155HG, it can be seen that with the increase of the T phase, except that the expression of T1 and T2 is similar, the overall trend is downward. The clinical correlation analysis of the stage is shown in **Figure4(C)**. MIR155HG was the most associated with risk, and its expression gradually decreased with the increase of stage. The clinical correlation analysis of N is shown in **Figure4(D)**.

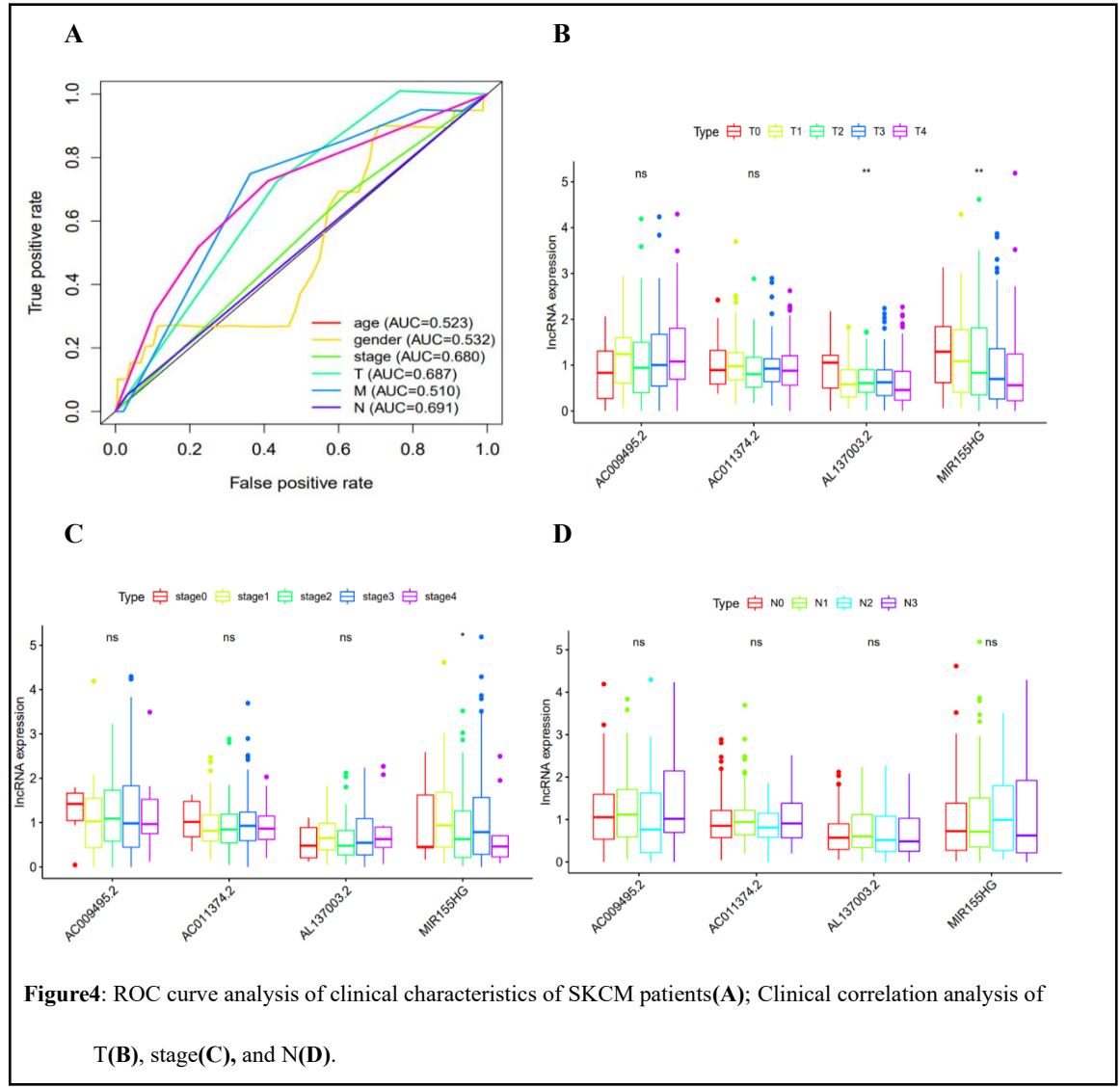

**Figure4**: ROC curve analysis of clinical characteristics of SKCM patients**(A)**; Clinical correlation analysis of T**(B)**, stage**(C),** and N**(D)**.

Surprisingly, it had the highest AUC value, but no lncRNA was significantly correlated with it. As can be seen from the figure, AC009495.2, AC011374.2, and MIR155HG are all expressed more with the increase of N, so we guess that they may just not reach the threshold, so there is no significant correlation of lncRNA.

**A comprehensive evaluation of predictive model performance and conventional clinical risk factors**

We compared the predictive performance of the 4 immune-related lncRNAs model with traditional clinical risk factors, including age, gender, stage, T, M and N. Univariate analysis showed that the age, stage, T, N, M and prediction performance of the 4 immune-related lncRNAs model were firmly related to the prognosis(**Figure5(A)**). Further multivariate analysis showed that T、N and 4 immune-related lncRNAs models could be used as independent prognostic factors(**(Figure5(B)**).

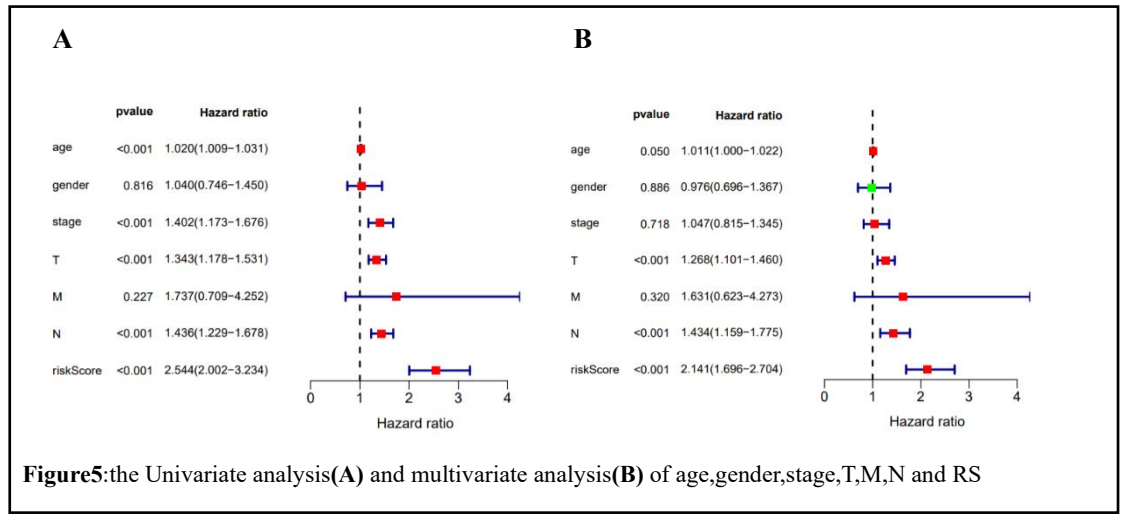

**Figure5**:the Univariate analysis**(A)** and multivariate analysis**(B)** of age,gender,stage,T,M,N and RS

**Visualization of expression differences between high-risk and low-risk patients**

We reduce the dimensions by Principal Component Analysis ( PCA) to visualize the data, and the results are shown in the figure5. According to the four scatter plots(**Figure5**), High-risk patients and low-risk patients can be well distinguished by lncRNAs of risk model, immune lncRNAs, immune genes, and all genes. Most notably, lncRNAs based on risk models identify high-risk patients from low-risk patients, which further proves the accuracy of the model.

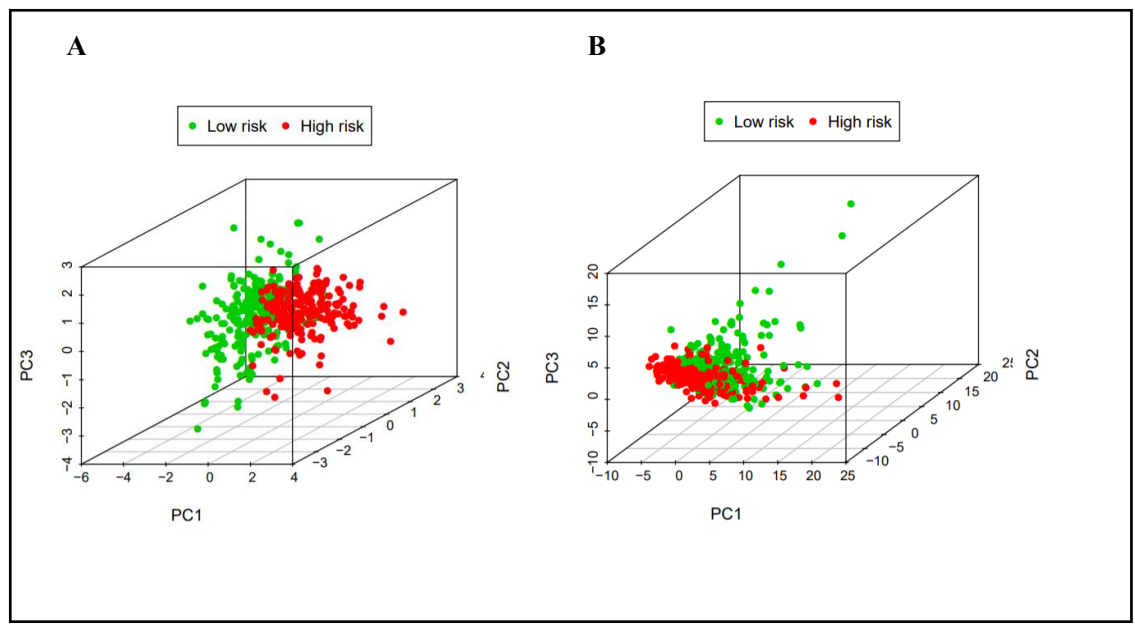

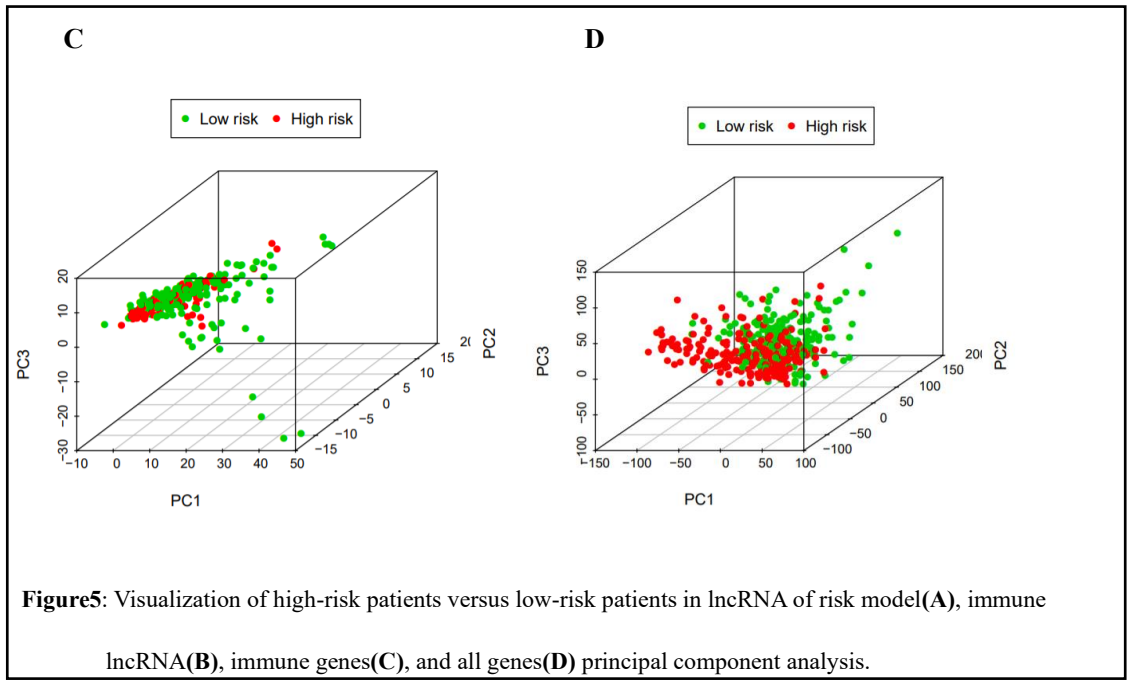

**Figure5**: Visualization of high-risk patients versus low-risk patients in lncRNA of risk model**(A)**, immune lncRNA**(B)**, immune genes**(C)**, and all genes**(D)** principal component analysis.

**Functional evaluation of lncRNA - associated mRNAs**

Based on SKCM-related lncRNA and mRNA expression data in the TCGA database, Pearson correlation analysis was used to analyze |COR | > 0.25 and P-value < 0.05 as for coexpression analysis. A total of 616 mRNAs were found to be closely correlated with seven lncRNA (**Figure 6**), and the function of lncRNA-related mRNAs was determined by using R org.Hs.eg.db package.

Results mainly included biological process (BP), molecular function (MF), and cell composition (CC). We selected the ten most significant enrichment results for analysis in three parts. The enrichment process of BP mainly includes T cell activation, regulation of lymphocyte activation, regulation of T cell activation, and leukocyte cell−cell adhesion, which are strictly related to the growth and reproduction of tumor cells. The enrichment characteristics of MF mainly include cytokine receptor activity, MHC protein binding, and cytokine binding. The 616 mRNAs were primarily enriched in 30 signaling pathways (**Figure6(B)**), including Cytokine−cytokine receptor interaction, Chemokine signaling pathway, and cell adhesion molecules (CAMs) and Human T−cell leukemia virus one infection. Most of these thirty pathways are related to the immune system of bettas, as is Th1 and Th2 cell differentiation and T cell receptor signaling pathway

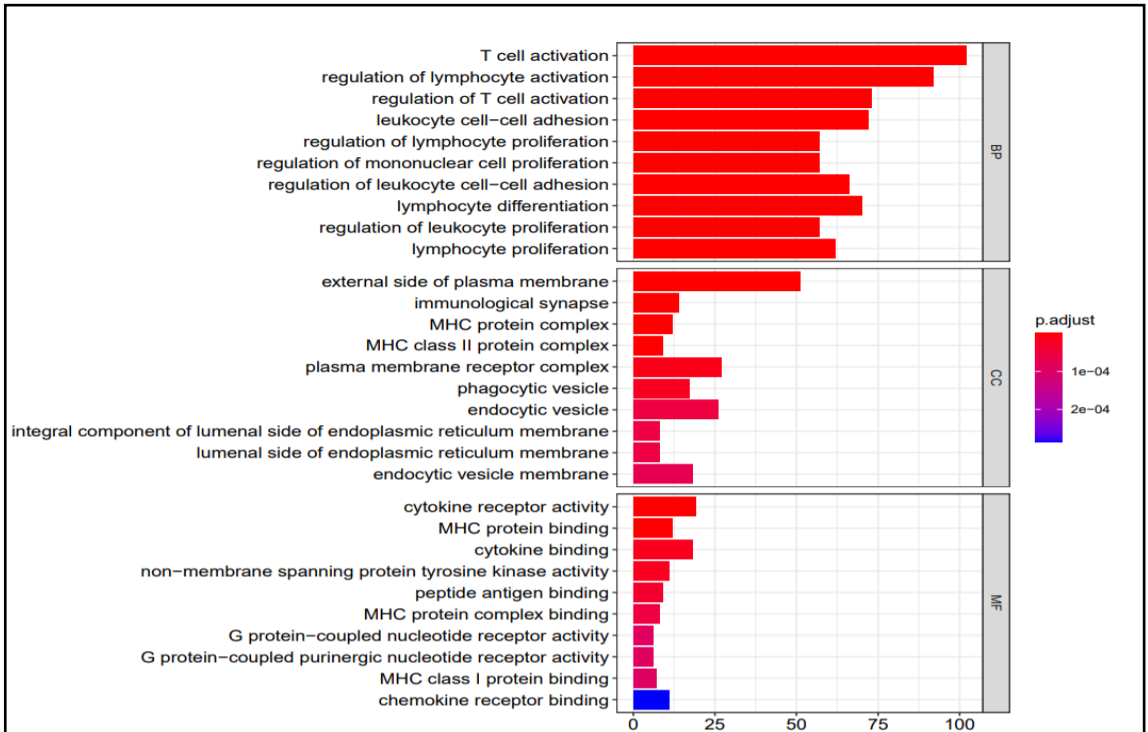

**Figure6(A)**: Enrichment analysis of biological processes, molecular function, and cellular component (P < 0.05)

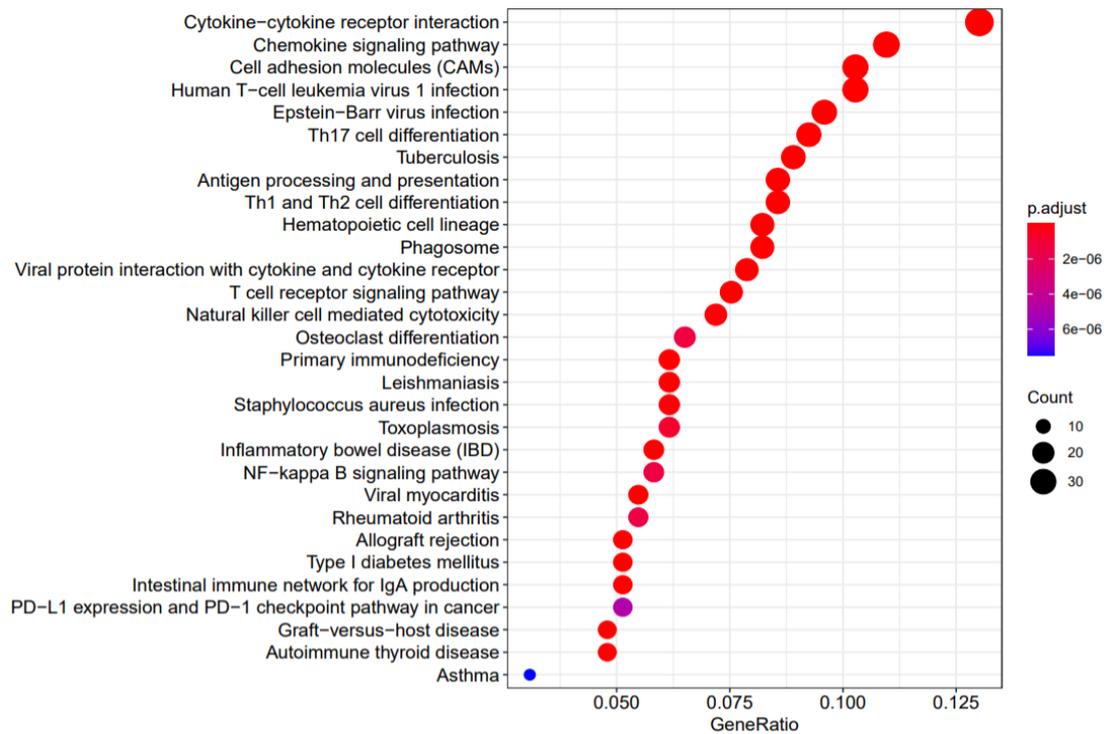

**Figure6(B):** Pathways enrichment map of lncRNA-related mRNAs

# DISCUSSION

SKCM remains one of the deadliest malignancies in the world[29]. Due to its complex molecular and cellular heterogeneity[30], the effectiveness of existing SKCM risk prediction models is not satisfactory[31]. Therefore, it is necessary to construct a new SKCM risk prediction model for clinical use to reduce mortality and improve the prognosis of SKCM. Clinicians should develop personalized treatment plans for SKCM patients, develop strategies for prevention and early detection of SKCM, track high-risk populations more frequently, conduct regular clinical examinations based on model predictions, and make an early diagnosis.

In this study, differential expressions of lncRNA and mRNA related to SKCM were obtained by high-throughput RNA sequencing and clinical data from SKCM patients in the TCGA database. Subsequently, we extracted immune-related lncRNA. Finally, univariate and multivariate Cox analyses were performed to establish a risk model for predicting the prognosis of SKCM. Four lncRNA (MIR155HG, AL137003.2, AC011374.2, AC009495.2) were used to construct the prediction model of SKCM prognostic risk. Applying the prediction model to the TCGA SKCM data set, SKCM patients can be divided into a high-risk group and low-risk group. The AUC value of the model is 0.722, indicating that the 4 immune-related lncRNAs model has excellent survival prediction performance. Of the four lncRNA, MIR155HG was found to be a marker of early cancer development[32]. MIR155HG is activated and upregulated by the MYB transcription factor[33], leading to the downregulation of many tumor suppressor genes. The expression of lncRNA MIR155 host gene MIR155HG was significantly correlated with the infiltration level of immune cells and immune molecules, especially with the appearance of immune checkpoint molecules such as programmed cell death protein 1 (PD-1), PD-1 ligand 1 (PD-L1), and cytotoxic T lymphocyte-associated antigen 4 (CTLA4) in most cancers[34]. Detection of clinical CHOL and hepatocellular carcinoma tissues confirmed that MIR155HG expression was positively correlated with CTLA4 and PD-L1 levels[35].

So far, studies on the remaining three lncRNA are not in-depth enough.AC009495 is a highly transcribed HERV gene[36]. HERV forms an essential part of the human genome, with about

98,000 ERV elements and fragments, accounting for 5-8%. Blast analysis showed that AC011374 was related to CTC-543d15[37], while CTC-543d15 was related to astrocytoma and rectal cancer[38].

The use of the TCGA database provides data for the prediction model of cancer survival. Compared with the previously constructed SKCM lncRNA prognosis model[39], the TCGA database has a large amount of patient sample data, complete clinical information, and full prognosis and survival data of SKCM patients. The ROC curve can be used to evaluate the specificity and sensitivity of the model (AUC >0.7 means the model has excellent responsiveness). The prognostic model of 4 immune-related lncRNAs established by us has the potential to predict the prognosis of SKCM patients with specificity and sensitivity. Besides, whether univariate cox-regression analysis or multivariate cox-regression analysis, the predictive effect of the 4 immune-related lncRNAs model constructed by us can evaluate the prognosis and further illustrate the evaluation value of the model. Besides, since the lncRNA used in this model have a predictive effect on the prediction of SKCM patients, also experimental studies can be carried out to explore the role of these lncRNAs in the pathogenesis of BRCA, providing new ideas and insights for treatment. However, the present study still has some limitations, and we attempted to verify the predictive performance of the 4 immune-related lncRNAs model in other large SKCM data sets. Unfortunately, due to the restrictions of SKCM clinical mutation information and patient prognosis information, we did not find a data set that met the validation requirements. Therefore, it is necessary to propose effective strategies, such as extending the follow-up time to verify the results and adopting a machine learning method to improve the accuracy of the model.

## Conclusion

We constructed a 4 immune-related lncRNAs model to predict the prognosis of SKCM patients reliably, and these lncRNAs may play a role in the carcinogenesis of SKCM. Further functional studies are needed to elucidate the molecular mechanisms behind the purpose of these lncRNAs in SKCM.

# Data availability declaration

This manuscript contains previously unpublished data. The name of the repository and accession number are not available.

Validation of a SKCM Risk Prediction Model Based on Self-assessed Risk Factors," *JAMA Dermatol,* vol. 152, no. 8, pp. 889-96, Aug 1, 2016.